\documentclass[aps,prc,superscriptaddress,twoside,twocolumn,nofootinbib,showpacs,floatfix]{revtex4-1}
\usepackage{amsmath,amssymb}
\usepackage{graphicx}
\usepackage{ulem}
\usepackage{subfigure}
\usepackage[colorlinks]{hyperref}
\usepackage[usenames,dvipsnames]{color}
\pdfpagewidth=\paperwidth
\pdfpageheight=\paperheight
\allowdisplaybreaks
\usepackage{newtxtext, newtxmath}
\usepackage{braket,bm}
\usepackage{multirow}
\usepackage{enumitem}
\usepackage[vcentermath]{youngtab}
\usepackage{appendix}
\usepackage{gensymb}

\begin{document}

\title{\boldmath Hidden charm tetraquark states in a diquark model}

\author{Pan-Pan Shi}
\affiliation{School of Nuclear Science and Technology, University of Chinese Academy of Sciences, Beijing 101408, China}
\affiliation{CAS Key Laboratory of Theoretical Physics, Institute of Theoretical Physics, Chinese Academy of Sciences, Beijing 100190, China}

\author{Fei Huang}
\affiliation{School of Nuclear Science and Technology, University of Chinese Academy of Sciences, Beijing 101408, China}

\author{Wen-Ling Wang}
\email[Email: ]{wangwenling@buaa.edu.cn}
\affiliation{School of Physics, Beihang University, Beijing 100191, China}

\date{\today}

\begin{abstract}
The purpose of the present study is to explore the mass spectrum of the hidden charm tetraquark states within a diquark model. Proposing that a tetraquark state is composed of a diquark and an antidiquark, the masses of all possible $[qc][\bar{q}\bar{c}]$, $[sc][\bar{s}\bar{c}]$, and $[qc][\bar{s}\bar{c}]$ $\left([sc][\bar{q}\bar{c}]\right)$ hidden charm tetraquark states are systematically calculated by use of an effective Hamiltonian, which contains color, spin, and flavor dependent interactions. Apart from the $X(3872)$, $Z(3900)$, $\chi_{c2}(3930)$, and $X(4350)$ which are taken as input to fix the model parameters, the calculated results support that the $\chi_{c0}(3860)$, $X(4020)$, $X(4050)$ are $[qc][\bar{q}\bar{c}]$ states with $I^GJ^{PC}=0^+0^{++}$, $1^+1^{+-}$, and $1^-2^{++}$, respectively, the $\chi_{c1}(4274)$ is an $[sc][\bar{s}\bar{c}]$ state with $I^GJ^{PC}=0^+1^{++}$, the $X(3940)$ is a $[qc][\bar{q}\bar{c}]$ state with $I^GJ^{PC}=1^-0^{++}$ or $1^-1^{++}$, the $Z_{cs}(3985)^-$ is an $[sc][\bar{q}\bar{c}]$ state with $J^{P}=0^{+}$ or $1^+$, and the $Z_{cs}(4000)^+$ and $Z_{cs}(4220)^+$ are $[qc][\bar{s}\bar{c}]$ states with $J^{P}=1^{+}$. Predictions for other possible tetraquark states are also given.
\end{abstract}



\maketitle

\section{Introduction}

Since the first $XYZ$ state $X(3872)$ with $I^GJ^{PC} = 0^+1^{++}$, mass $M = 3872 \pm 0.6$ MeV, and width $\Gamma \leq 2.3$ MeV was accidentally discovered by the Belle Collaboration \cite{Choi2003} in the reaction $B \rightarrow KX$, $X \rightarrow \pi^+ \pi^- J/\psi$, many other charmoniumlike exotics were successively reported in experiments. For example, the $Z_c(3900)$ was found in 2013 by the BESIII Collaboration with $M = 3899.0 \pm 3.6$ MeV and $\Gamma = 46 \pm 10$ MeV in the reaction $e^+ e^-\rightarrow \pi^+ \pi^- J/\psi$ \cite{Ablikim2013}, and by the Belle Collaboration with $M = 3894.5 \pm 6.6$ MeV and $\Gamma = 63 \pm 24$ MeV in the reaction $e^+ e^-\rightarrow \gamma \pi^+ \pi^- J/\psi$ \cite{Liu2013}; the $\chi_{c1}(4274)$ was first found by the CDF and CMS Collaborations \cite{Chatrchyan2014,Aaltonen2017}, and later was confirmed by the LHCb Collaboration with $M = 4273.3  \pm 8.3$ MeV and $\Gamma = 56.2 \pm 10.9$ MeV \cite{Aaij2017}; the $\chi_{c0}(3860)$ was observed by the Belle Collaboration in the process $e^+e^-\rightarrow  J/\psi D\bar{D}$ with $M = 3862^{+26+40}_{-32-13}$ MeV and $\Gamma = 201^{+154+88}_{-67-82}$ MeV \cite{Chilikin2017}. Very recently, the BESIII Collaboration reported the $Z_{cs}(3985)^-$ with $M = 3982.5^{+1.8}_{-2.6} \pm 2.1$ MeV and $\Gamma = 12.8^{+5.3}_{-4.4} \pm 3.0$ MeV, which is a first candidate of the charged hidden-charm tetraquark with strangeness \cite{Ablikim:2020hsk}; the LHCb Collaboration observed the $Z_{cs}(4000)^+$ with $M = 4003 \pm 6^{+4}_{-14}$ MeV and $\Gamma = 131 \pm 15 \pm 26$ MeV, and $Z_{cs}(4220)^+$ with $M = 4216 \pm 24^{+43}_{-30}$ MeV and $\Gamma = 233 \pm 52^{+97}_{-73}$ MeV \cite{Aaij:2021ivw}. More $XYZ$ states gradually discovered in experiments in recent years are listed in Ref.~\cite{Lebed2017}. 

Theoretically, several approaches were used to understand those $XYZ$ states and to predict new exotics. In particular, many $XYZ$ sates were explained as molecules, kinematical effects, or hybrids \cite{Bugg2008,Berwein2015}. See Refs.~\cite{Guo2018,Liu2019} for recent reviews. Besides those models and phenomenological explanations, the $XYZ$ states were also widely explored in diquark models \cite{Maiani2005,Hogaasen2006,Abud2010,Patel2014,Maiani2014,Cleven2015,Zhu2016,Kim2016,Wu2016,Maiani2016,Lebed2016,Giron2019,Jin:2020yjn,Wang:2020iqt,Lebed:2017min,Ali:2017wsf, Giron:2020fvd, Giron:2020fvd,Ozdem:2021yvo,Giron:2020qpb}. 

The $X(3872)$ was the first $XYZ$ state which has been discussed as a tetraquark state composed of a diquark $[qc]$ and an antidiquark $[\bar{q}\bar{c}]$ with $J^{P}=1^{+}$ \cite{Maiani2005,Hogaasen2006}. Later, efforts were also made in describing the $Z_c(3900)$ and $X(3915)$ as tetraquark states composed of a pair of diquark and antidiquark with $J^{P}=1^{+}$ and $0^+$, respectively \cite{Hogaasen2006,Patel2014,Maiani2014,Lebed:2017min}. In Refs.~\cite{Abud2010,Wu2016,Maiani2016,Lebed2016}, the $Y(4140)$ was assigned as a tetraquark state composed of a diquark $[sc]$ and an antidiquark $[\bar{s}\bar{c}]$ with $J^{PC}=1^{++}$, and meanwhile, the $X(4350)$ was explained as an $[sc][\bar{s}\bar{c}]$ tetraquark state with $J^{PC}=0^{++}$ or $2^{++}$. For $\chi_{c1}$ states, as both $\chi_{c1}(4140)$ and $\chi_{c1}(4274)$ have the same quantum numbers, $I^GJ^{PC}=0^+1^{++}$, it is still not clear which one should be identified as the only theoretical $S$-wave $[sc][\bar{s}\bar{c}]$ tetraquark state with $I^GJ^{PC}=0^+1^{++}$ \cite{Zhu2016,Maiani2016,Wu2016,Giron:2020qpb}, an analog of the $X(3872)$ which was considered as a $[qc][\bar{q}\bar{c}]$ tetraquark state with $I^GJ^{PC}=0^+1^{++}$ \cite{Maiani2005,Hogaasen2006}. Besides, the $Z_{cs}(3985)^-$ was analyzed as a candidate of the charged hidden-charm tetraquark with strangeness composed of a diquark $[sc]$ and an antidiquark $[\bar{q}\bar{c}]$ with $J^P=1^+$ or $0^+$ \cite{Jin:2020yjn,Wang:2020iqt,Ozdem:2021yvo}.

In most of the existing diquark models for tetraquark states, the flavor dependent interactions are not included in the effective model Hamiltonian, which makes these models fail to describe the mass splitting among different isospin multiplets of tetraquark states with the same spin, parity, and diquark constituents. In Ref.~\cite{Giron2019}, the one-pion exchange potential was considered to distinguish the isospin-triplet and isospin-singlet tetraquark states; nevertheless, the color-spin interactions between diquarks were ignored in that model. 

In the present work, we use an effective Hamiltonian which contains explicitly the color-spin, spin, flavor-spin, and flavor dependent interactions to investigate the mass spectrum of the hidden charm tetraquark states composed of diquarks and antidiquarks. We systematically calculate the masses of all possible $[qc][\bar{q}\bar{c}]$, $[sc][\bar{s}\bar{c}]$, and $[qc][\bar{s}\bar{c}]$ $\left([sc][\bar{q}\bar{c}]\right)$ hidden charm tetraquark states, and discuss which of the observed $XYZ$ states can be accommodated in a diquark scenario and which cannot. We also give predictions for the possible tetraquark states which have not yet been observed so far.

The paper is organized as follows. In Sec.~\ref{Sect:wave function}, we construct the wave functions for the hidden charm tetraquark states. In Sec.~\ref{Sect:Mass_formula}, we present the effective Hamiltonian employed in the present work and give the mass formula for the tetraquark states considered. In Sec.~\ref{Sect:Numerical_result}, we show the results of the calculated masses of the hidden charm tetraquark states and give discussions for these theoretical results. Finally, in Sec.~\ref{Sect:Summary}, we give a brief summary.

\section{Wave functions of the hidden charm tetraquark states}  \label{Sect:wave function}

In color space, each quark (antiquark) belongs to a color ${\bm 3}_c$ (${\bar{\bm 3}}_c$) representation, two quarks can stay in color ${\bar{\bm 3}}_c$ or ${\bm 6}_c$ configurations, and two antiquarks can stay in color ${{\bm 3}}_c$ or ${\bar{\bm 6}}_c$ configurations. It is well known that the interactions arising from QCD are attractive between a pair of quarks in color ${\bar{\bm 3}}_c$ state or a pair of antiquarks in color ${{\bm 3}}_c$ state \cite{Maiani2005}. Thus, the color wave function for a tetraquark state composed of a pair of diquark and antidiquark can be constructed straightforwardly as $\left[\left({\bm 3}_c {\bm 3}_c\right)_{{\bar{\bm 3}}_c} ({\bar{\bm 3}}_c {\bar{\bm 3}}_c)_{{\bm 3}_c}\right]_{{\bm 1}_c}$.

In flavor space, a hidden charm tetraquark state composed of a pair of diquark and antidiquark can be denoted as $[q'c][\bar{q}'\bar{c}]$, where $q'$ represents one of the three light quarks, $u, d$, and $s$, in the flavor SU(3) case. The light quark $q'$ and light antiquark $\bar{q}'$ combine into an SU(3) octet multiplet or an SU(3) singlet multiplet:
\begin{align}
{\yng(1)}_{\,{\bf{3}}_f}  ~\otimes~  {\yng(1,1)}_{\,{\bar{\bf{3}}}_f}  ~\simeq~  {\yng(2,1)}_{\,{\bf{8}}_f}  ~\oplus~ {\yng(1,1,1)}_{\,{\bf{1}}_f}.\label{Eq:flavor_function}
\end{align}
Thus, similar to the $q'\bar{q}'$-meson octet and singlet multiplets, the hidden charm tetraquark states can decompose into a flavor SU(3) octet multiplet
\begin{align}
{M^a}_b = 
\begin{pmatrix}
 \frac{Z^0_c}{\sqrt{2}} + \frac{X}{\sqrt{6}}    &  Z_c^+    &  Z^+_{cs}  \\
  Z^-_c      & -\frac{Z^0_c}{\sqrt{2}} + \frac{X}{\sqrt{6}}   &   Z^0_{cs}  \\
  Z^-_{cs}  & \bar{Z}^0_{cs}  & -\frac{2X}{\sqrt{6}}
 \end{pmatrix},
\label{Eq:flavor_octet}
\end{align}
and a flavor SU(3) singlet $X'$. Explicitly, the flavor wave functions for the states denoted by the symbols $Z^0_c$, $X$, $Z_c^+$, $Z^+_{cs}$, $Z^-_c$, $Z^0_{cs}$, $Z^-_{cs}$, $\bar{Z}^0_{cs}$, and $X'$ can be written as \cite{Zhu2016}
\begin{align}
& Z^-_c = [dc][\bar{u}\bar{c}]\,,  \quad \qquad Z_c^+ = [uc] [\bar{d}\bar{c}]\,,  \label{Eq:flavor_wave_function_0}  \\
& Z^0_c = \frac{1}{\sqrt{2}}\left([uc][\bar{u}\bar{c}]-[dc][\bar{d}\bar{c}]\right),  \\
& Z^0_{cs} = [dc][\bar{s}\bar{c}]\,,  \quad \qquad  Z^+_{cs} = [uc][\bar{s}\bar{c}]\,,  \label{Eq:flavor_wave_function_5} \\[3pt]
& Z^-_{cs} = [sc][\bar{u}\bar{c}]\,,  \quad \qquad  {\bar Z}^0_{cs} = [sc][\bar{d}\bar{c}]\,,  \label{Eq:flavor_wave_function_6}  \\
& X = \frac{1}{\sqrt{6}}\left([uc][\bar{u}\bar{c}] + [dc][\bar{d}\bar{c}] - 2[sc][\bar{s}\bar{c}]\right),  \\
& X' = \frac{1}{\sqrt{3}}\left([uc][\bar{u}\bar{c}] + [dc][\bar{d}\bar{c}] + [sc][\bar{s}\bar{c}]\right).
\label{Eq:flavor_wave_function}
\end{align}
The symmetry states $X$ from the flavor SU(3) octet and $X'$ from the flavor SU(3) singlet have the same  isospin $0$. They can couple together to result in the physically observed states $\bar{X}$ and $\bar{X}'$:
\begin{align}
 \begin{pmatrix}
 \bar{X}'   \\ \bar{X}       
 \end{pmatrix}
 =
  \begin{pmatrix}
 \cos\theta  &  -\sin\theta   \\
 \sin\theta  &  \cos\theta   
 \end{pmatrix}
 \begin{pmatrix}
   X'  \\   X    
 \end{pmatrix},
 \label{Eq:flavor_mixed}
 \end{align}
with $\theta$ being the mixing angle. For simplicity, an ``ideal mixing'' is chosen, i.e. $\theta \approx 54.732\degree$, which gives
\begin{align}
&\bar{X} = \frac{1}{\sqrt{2}}\left([uc][\bar{u}\bar{c}] + [dc][\bar{d}\bar{c}]\right), \label{Eq:ideal_mixing_0}
\\
&\bar{X}' = [sc][\bar{s}\bar{c}].    \label{Eq:ideal_mixing}
\end{align}
 In short, the above-mentioned flavor wave functions for hidden charm tetraquark states composed of diquarks and antidiquarks could be divided to three types of configurations, i.e. $[qc][\bar{q}\bar{c}]$, $[sc][\bar{s}\bar{c}]$, and $[qc][\bar{s}\bar{c}]$ $\left([sc][\bar{q}\bar{c}]\right)$, where $q$ denotes $u$ or $d$ quark, and $\bar{q}$ denotes $\bar{u}$ or $\bar{d}$ quark. Note that here $q\bar{q}$ could be one of $u\bar{u}$, $u\bar{d}$, $d\bar{u}$, and $d\bar{d}$. In the rest of this paper, we refer to the diquarks $[qc]$ and $[sc]$ as ${\cal D}$ and ${\cal D}_s$, the antidiquarks $[\bar{q}\bar{c}]$ and $[\bar{s}\bar{c}]$ as ${\bar{\cal D}}$ and ${\bar{\cal D}}_s$, respectively, for the sake of simplicity.

In spin space, the wave functions for the three flavor types of configurations, ${\cal D}{\bar{\cal D}}$, ${\cal D}_s{\bar{\cal D}}_s$, and ${\cal D}{\bar{\cal D}}_s$ (${\cal D}_s{\bar{\cal D}}$), can be constructed as follows. We mention that no orbital excitation inside a diquark or between a pair of diquark-antidiquark is considered in the present work.

\subsection{Spin wave functions for ${\cal D}{\bar{\cal D}}$ and ${\cal D}_s{\bar{\cal D}}_s$}\label{SubSect:Spin_ccss}

As can be seen from Eqs.~(\ref{Eq:flavor_wave_function_0})$-$(\ref{Eq:ideal_mixing}), the isospin triplet $Z^-_c$, $Z^0_c$, and $Z^+_c$, and isospin singlet $\bar{X}$ are tetraquark states that have the ${\cal D}{\bar{\cal D}}$ type configurations, and the isospin singlet $\bar{X}'$ is the tetraquark state that has the ${\cal D}_s{\bar{\cal D}}_s$ type configuration. Their spin wave functions can be constructed as \cite{Maiani2014,Lebed:2017min,Giron2019}
\begin{align}
&J^{PC}=0^{++}: \quad  \Ket{0, 0}_0, \quad \Ket{1, 1}_0;  \label{Eq:Spin_wave_qc_qc_0} \\
&J^{PC}=1^{++}:  \quad  \frac{1}{\sqrt{2}}\left(\Ket{1, 0}_1 + \Ket{0, 1}_1\right); \label{Eq:Spin_wave_qc_qc_1}  \\
&J^{PC}=1^{+-}:  \quad  \frac{1}{\sqrt{2}}\left(\Ket{1, 0}_1 - \Ket{0, 1}_1\right),  \quad \Ket{1, 1}_1;  \label{Eq:Spin_wave_qc_qc_2}  \\
&J^{PC}=2^{++}:  \quad  \Ket{1, 1}_2.    \label{Eq:Spin_wave_qc_qc}
\end{align}
Here each ket denotes $\Ket{S_{\cal D}, S_{\bar{\cal D}}}_{S_{{\cal D}{\bar{\cal D}}}}$ or $\Ket{S_{{\cal D}_s}, S_{{\bar{\cal D}}_s}}_{S_{{\cal D}_s{\bar{\cal D}}_s}}$, with $S$ being the spin of the diquark (antidiquark) or a pair of diquark and antidiquark indicated by the corresponding subscript. Note that for ${\cal D}_s{\bar{\cal D}}_s$ states the isospin is $I=0$ and the $G$ parity is given by $G=C(-1)^I=C$.

\subsection{Spin wave functions for ${{\cal D}{\bar{\cal D}}_s}$ or ${{\cal D}_s{\bar{\cal D}}}$}

One sees from Eqs.~(\ref{Eq:flavor_wave_function_0})$-$(\ref{Eq:ideal_mixing}), the isospin doublet $Z^0_{cs}$ and $Z^+_{cs}$ are tetraquark states that have the ${\cal D}{\bar{\cal D}_s}$ type configuration, and the isospin doublet $Z^-_{cs}$ and ${\bar Z}^0_{cs}$ are tetraquark states that have the ${\cal D}_s{\bar{\cal D}}$ type configuration. Their spin wave functions can be constructed as
\begin{align}
&J^{P}=0^{+}:  \quad \Ket{0, 0}_0, \quad \Ket{1, 1}_0;  \label{Eq:Spin_wave_qc_sc_0} \\
&J^{P}=1^{+}:   \quad \Ket{1, 0}_1, \quad \Ket{0, 1}_1, \quad \Ket{1, 1}_1; \label{Eq:Spin_wave_qc_sc_1} \\
&J^{P}=2^{+}:  \quad \Ket{1, 1}_2.  \label{Eq:Spin_wave_qc_sc}
\end{align}
Here each ket denotes $\Ket{S_{\cal D}, S_{\bar{\cal D}_s}}_{S_{{\cal D}{\bar{\cal D}_s}}}$ or $\Ket{S_{{\cal D}_s}, S_{{\bar{\cal D}}}}_{S_{{\cal D}_s{\bar{\cal D}}}}$, with $S$ being the spin of the diquark (antidiquark) or a pair of diquark and antidiquark indicated by the corresponding subscript. Note that both the ${{\cal D}{\bar{\cal D}}_s}$ and ${{\cal D}_s{\bar{\cal D}}}$ configuration states are constructed to have particular strangeness quantum numbers [c.f. Eqs.~(\ref{Eq:flavor_wave_function_5}) and (\ref{Eq:flavor_wave_function_6})], and consequently, they are not eigenstates of the charge conjugation operator.

\section{Mass formulas for the hidden charm tetraquark states} \label{Sect:Mass_formula}

\subsection{The effective Hamiltonian}

The phenomenological Hamiltonian of a diquark model is usually parametrized as a sum of the diquark masses and an effective potential composed of color-spin and color-electric interaction terms \cite{Maiani2005,Zhu2016,Shi2019} which are inspired by one-gluon exchange potential and instant-induced interaction\cite{Kim2016}. Such a potential can be written explicitly as \cite{Shi2019}
\begin{align}
V_{cs} = 2\sum_{i>j} \left[ \alpha_{ij} \left( {\bm \lambda}^c_i \cdot {\bm \lambda}^c_j {\bm S}_i \cdot {\bm S}_j \right) + \frac{\beta}{m_im_j} \left( {\bm \lambda}^c_i \cdot {\bm \lambda}^c_j \right) \right],
\label{Eq:Hamiltonian_0}
\end{align}
where the parameter $\alpha_{ij}$ and the masses $m_i$ and $m_j$ depend on the flavor of constituents $i$ and $j$, while the parameter $\beta$ is flavor independent. ${\bm \lambda}^c$ represents the Gell-Mann matrix for the color SU(3) group. In order to describe the mass splits of isospin multiplets of the considered tetraquark states, we introduce, analogously to the color dependent potential in Eq.~(\ref{Eq:Hamiltonian_0}), the following flavor dependent potential:
\begin{align}
V_{fs} = 2\sum_{i>j} \left[ \frac{\gamma}{m_{i}m_{j}} \left( {\bm \lambda}^f_{i} \cdot {\bm \lambda}^f_{j} {\bm S}_{i} \cdot {\bm S}_{j} \right) + \frac{\rho}{m_{i}m_{j}} \left( {\bm \lambda}^f_{i} \cdot {\bm \lambda}^f_{j} \right) \right],
\label{Eq:Hamiltonian_1}
\end{align}
where the parameters $\gamma$ and $\rho$ are both flavor independent, and ${\bm \lambda}^f$ represents the Gell-Mann matrix for the flavor SU(3) group. Note that in Eqs.~(\ref{Eq:Hamiltonian_0}) and (\ref{Eq:Hamiltonian_1}), the summations are performed over all pairs of quarks (antiquarks), either inside a diquark (antidiquark) or between a diquark and an antidiquark for a tetraquark state. The effective Hamiltonian of the model is then written as
\begin{align}
H= \sum_n M_n+V_{cs}+V_{fs},
\label{Eq:Hamiltonian}
\end{align}
with $M_n$ being the effective mass of the $n$th constituent which includes also those effects not accounted for by the above-mentioned interactions. 

The color matrix elements for tetraquark states composed of diquark-antidiquark are
\begin{align}
\Braket{{\bm \lambda}^c_i \cdot {\bm \lambda}^c_j} = \left\{ \begin{array}{lll} -8/3, && \left(i, j {\rm ~ in ~ the ~ same ~ diquark}\right) \\[2pt] -4/3. && \left({\rm others}\right) \end{array} \right.
 \label{Eq:color_matrix}
\end{align}

The flavor matrix elements are  
\begin{align}
\Braket{{\bm{\lambda}}^f_{i} \cdot {\bm{\lambda}}^f_{j}} = 2/3,
\label{Eq:flavor_octet_matrix}
\end{align}
for tetraquark states $Z^-_c$, $Z_c^0$, $Z^+_c$, $Z^0_{cs}$, $Z^+_{cs}$, $Z^-_{cs}$, and $\bar{Z}^0_{cs}$ defined in Eqs.~(\ref{Eq:flavor_wave_function_0})$-$(\ref{Eq:flavor_wave_function_6}), and are 
\begin{align}
& \Braket{{\bm{\lambda}}^f_{i} \cdot {\bm{\lambda}}^f_{j} }_{\bar{X}} = -10/3,  \\
& \Braket{{\bm{\lambda}}^f_{i} \cdot {\bm{\lambda}}^f_{j} }_{\bar{X}'} = -4/3,
\label{Eq:flavor_singlet_matrix}
\end{align}
for tetraquark states $\bar{X}$ and $\bar{X}'$ defined in Eqs.~(\ref{Eq:ideal_mixing_0})$-$(\ref{Eq:ideal_mixing}). 

The spin matrix elements can be calculated directly from the spin wave functions constructed in Eqs.~(\ref{Eq:Spin_wave_qc_qc_0})$-$(\ref{Eq:Spin_wave_qc_sc}). The details of the calculation are given in Appendix~\ref{Sect:spin_matrix}, and the results are listed in Table~\ref{Tab:Spin_matrix}.

\begin{table}[tb]
\caption{\label{Tab:Spin_matrix} Spin matrix elements for tetraquark states composed of diquark-antidiquark pairs. The spin states are denoted by $\Ket{S_{12},S_{{\bar 3}{\bar 4}}}_{S_{12{\bar 3}{\bar 4}}}$.}
\renewcommand{\arraystretch}{1.2}
\begin{tabular*}{\columnwidth}{@{\extracolsep\fill}lcccccc}
\hline\hline                                           
          & ${\bm S}_1\cdot {\bm S}_2$     & ${\bm S}_1\cdot {\bm S}_{\bar 3}$   & ${\bm S}_1\cdot {\bm S}_{\bar 4}$   &  ${\bm S}_2\cdot {\bm S}_{\bar 3}$   & ${\bm S}_2\cdot {\bm S}_{\bar 4}$   &  ${\bm S}_{\bar 3}\cdot {\bm S}_{\bar 4}$ \\                                               
\hline
$\Braket{0, 0 | {\bm S}_i \cdot {\bm S}_j | 0, 0}_0$   & $-3/4$   & $0$     & $0$      &  $0$       & $0$     & $-3/4$   \\
$\Braket{1, 0 | {\bm S}_i \cdot {\bm S}_j | 1, 0}_1$    & $1/4$     & $0$      & $0$      &  $0$       & $0$     & $-3/4$   \\
$\Braket{0, 1 | {\bm S}_i \cdot {\bm S}_j | 0, 1}_1$    & $-3/4$   & $0$      & $0$      &  $0$       & $0$     & $1/4$      \\
$\Braket{1, 1 | {\bm S}_i \cdot {\bm S}_j | 1, 1}_0$    & $1/4$     & $-1/2$  & $-1/2$  &  $-1/2$  & $-1/2$  & $1/4$    \\
$\Braket{1, 1 | {\bm S}_i \cdot {\bm S}_j | 1, 1}_1$    & $1/4$     & $-1/4$   & $-1/4$  &  $-1/4$  & $-1/4$  & $1/4$     \\
$\Braket{1, 1 | {\bm S}_i \cdot {\bm S}_j | 1, 1}_2$   & $1/4$     & $1/4$     & $1/4$    &  $1/4$    & $1/4$    & $1/4$   \\
$\Braket{0, 0 | {\bm S}_i \cdot {\bm S}_j | 1, 1}_0$ & $0$  & $-\sqrt{3}/4$  & $\sqrt{3}/4$  &  $\sqrt{3}/4$  & $-\sqrt{3}/4$  & $0$   \\
$\Braket{1, 0 | {\bm S}_i \cdot {\bm S}_j | 0, 1}_1$  & $0$ & $1/4$  & $-1/4$  & $-1/4$ &  $1/4$  & $0$  \\
$\Braket{0, 1 | {\bm S}_i \cdot {\bm S}_j | 1, 1}_1$   & $0$ & $-\sqrt{2}/4$  & $-\sqrt{2}/4$   &  $\sqrt{2}/4$  & $\sqrt{2}/4$  & $0$   \\
$\Braket{1, 0 | {\bm S}_i \cdot {\bm S}_j | 1, 1}_1$  & $0$ & $\sqrt{2}/4$  & $-\sqrt{2}/4$   &  $\sqrt{2}/4$  & $-\sqrt{2}/4$   & $0$  \\[2pt]
\hline\hline
\end{tabular*}
\end{table}

\subsection{Mass formulas for ${\cal D}{\bar{\cal D}}$}  \label{sec:mass_DDbar}

By use of the effective Hamiltonian in Eq.~(\ref{Eq:Hamiltonian}), the color and flavor matrix elements in Eqs.~(\ref{Eq:color_matrix})$-$(\ref{Eq:flavor_singlet_matrix}), and the spin matrix elements listed in Table~\ref{Tab:Spin_matrix}, the masses of hidden charm tetraquark states ${\cal D}{\bar{\cal D}}$, ${\cal D}_s{\bar{\cal D}}_s$, and ${\cal D}{\bar{\cal D}}_s$ (${\cal D}_s{\bar{\cal D}}$) can be calculated straightforwardly. 

For ${\cal D}{\bar{\cal D}}$ states with $I^G J^{PC}=0^+ 0^{++}$ and $I^G J^{PC}=1^- 0^{++}$, whose spin wave functions are defined in Eq.~(\ref{Eq:Spin_wave_qc_qc_0}), the masses are, respectively, determined by the following mass matrices:
\begin{align}
\begin{pmatrix}
 M_{(00)0,0}    & M'_{(00-11)0,0} \\
 M'_{(00-11)0,0}  & M_{(11)0,0}                                                    
\end{pmatrix},  \label{Eq:mass_qc_qc_0_0}    \\[3pt]
\begin{pmatrix}
 M_{(00)0,1}    & M'_{(00-11)0,1} \\
 M'_{(00-11)0,1}  & M_{(11)0,1}                                                    
\end{pmatrix},   \label{Eq:mass_qc_qc_0_1}
\end{align}
where the subscripts of the diagonal matrix elements $M$ denote $\left(S_{\cal{D}}S_{\bar{\cal{D}}}\right)S,I$, and the subscripts of the transition matrix elements $M'$ denote $\left(S_{\cal{D}}S_{\bar{\cal{D}}}-S'_{\cal{D}}S'_{\bar{\cal{D}}}\right)S,I$, with $S$ and $I$ being the total spin and isospin for the considered tetraquark states. Explicitly, one has
\begin{align}
M_{(00)0,0} = &\; 2M_{qc}+8\alpha_{qc}-\frac{8\beta}{3m_q^2}-\frac{16\beta}{m_qm_c}-\frac{8\beta}{3m_c^2} -\frac{20\rho}{3m_q^2},  \\
M_{(11)0,0} = &\; 2M_{qc}+\frac{4}{3}\alpha_{qq}+\frac{4}{3}\alpha_{cc}-\frac{8\beta}{3m_q^2}-\frac{16\beta}{m_qm_c} -\frac{8\beta}{3m_c^2}  \nonumber \\
                   &\; -\frac{20\rho}{3m_q^2}+\frac{10\gamma}{3m_q^2},   \\
M'_{(00-11)0,0} = &\; \frac{2}{\sqrt{3}}\alpha_{qq}-\frac{4}{\sqrt{3}}\alpha_{qc}+\frac{2}{\sqrt{3}}\alpha_{cc}+\frac{5\gamma}{\sqrt{3}m_q^2},  \\
M_{(00)0,1} = &\; 2M_{qc}+8\alpha_{qc}-\frac{8\beta}{3m_q^2}-\frac{16\beta}{m_qm_c}-\frac{8\beta}{3m_c^2}  +\frac{4\rho}{3m_q^2},  \\
M_{(11)0,1} = &\; 2M_{qc}+\frac{4}{3}\alpha_{qq}+\frac{4}{3}\alpha_{cc}-\frac{8\beta}{3m_q^2}-\frac{16\beta}{m_qm_c} -\frac{8\beta}{3m_c^2}  \nonumber \\
                  &\; +\frac{4\rho}{3m_q^2}-\frac{2\gamma}{3m_q^2},   \\
M'_{(00-11)0,1} = &\; \frac{2}{\sqrt{3}}\alpha_{qq}-\frac{4}{\sqrt{3}}\alpha_{qc}+\frac{2}{\sqrt{3}}\alpha_{cc}-\frac{\gamma}{\sqrt{3}m_q^2}.
\end{align}
Similarly, for ${\cal D}{\bar{\cal D}}$ states with $I^G J^{PC}=0^- 1^{+-}$ and $I^G J^{PC}=1^+ 1^{+-}$, whose spin wave functions are defined in Eq.~(\ref{Eq:Spin_wave_qc_qc_2}), the masses are, respectively, determined by the following mass matrices:
\begin{align}
\begin{pmatrix}
 M_{(01)1,0}    & M'_{(01-11)1,0} \\
 M'_{(01-11)1,0}  & M_{(11)1,0}                                                    
\end{pmatrix},  \label{Eq:mass_qc_qc_1_0}    \\[3pt]
\begin{pmatrix}
 M_{(01)1,1}    & M'_{(01-11)1,1} \\
 M'_{(01-11)1,1}  & M_{(11)1,1}                                                    
\end{pmatrix},   \label{Eq:mass_qc_qc_1_1}
\end{align}
where 
\begin{align}
M_{(01)1,0} = &\; 2M_{qc}+\frac{2}{3}\alpha_{qq}+\frac{4}{3}\alpha_{qc}+\frac{2}{3}\alpha_{cc}-\frac{8\beta}{3m_q^2} -\frac{16\beta}{m_qm_c} \nonumber \\
                   &\; -\frac{8\beta}{3m_c^2}-\frac{20\rho}{3m_q^2}+\frac{5\gamma}{3m_q^2}, \\
M_{(11)1,0} = &\; 2M_{qc}+\frac{2}{3}\alpha_{qq}-\frac{4}{3}\alpha_{qc}+\frac{2}{3}\alpha_{cc}-\frac{8\beta}{3m_q^2} -\frac{16\beta}{m_qm_c} \nonumber \\
                   &\; -\frac{8\beta}{3m_c^2}-\frac{20\rho}{3m_q^2}+\frac{5\gamma}{3m_q^2},  \\
M'_{(01-11)1,0} = &\; -\frac{4}{3}\alpha_{qq}+\frac{4}{3}\alpha_{cc}-\frac{10\gamma}{3m_q^2},  \\
M_{(01)1,1} = &\; 2M_{qc}+\frac{2}{3}\alpha_{qq}+\frac{4}{3}\alpha_{qc}+\frac{2}{3}\alpha_{cc}-\frac{8\beta}{3m_q^2} -\frac{16\beta}{m_qm_c} \nonumber\\
                  &\; -\frac{8\beta}{3m_c^2}+\frac{4\rho}{3m_q^2}-\frac{\gamma}{3m_q^2},  \\
M_{(11)1,1} = &\; 2M_{qc}+\frac{2}{3}\alpha_{qq}-\frac{4}{3}\alpha_{qc}+\frac{2}{3}\alpha_{cc}-\frac{8\beta}{3m_q^2} -\frac{16\beta}{m_qm_c} \nonumber\\
                 &\; -\frac{8\beta}{3m_c^2}+\frac{4\rho}{3m_q^2}-\frac{\gamma}{3m_q^2},  \\
M'_{(01-11)1,1} = &\; -\frac{4}{3}\alpha_{qq}+\frac{4}{3}\alpha_{cc}+\frac{2\gamma}{3m_q^2}.
\end{align}
For ${\cal D}{\bar{\cal D}}$ states with $I^G J^{PC}=0^+1^{++}$ and $I^G J^{PC}=1^- 1^{++}$, whose spin wave functions are defined in Eq.~(\ref{Eq:Spin_wave_qc_qc_1}), the masses are, respectively, given by
\begin{align}
M_{(10)1,0} = &\; 2M_{qc}-\frac{2}{3}\alpha_{qq}+4\alpha_{qc}-\frac{2}{3}\alpha_{cc}-\frac{8\beta}{3m_q^2} -\frac{16\beta}{m_qm_c} \nonumber\\
                   &\; -\frac{8\beta}{3m_c^2}-\frac{20\rho}{3m_q^2}+\frac{5\gamma}{3m_q^2},   \\
M_{(10)1,1} = &\; 2M_{qc}-\frac{2}{3}\alpha_{qq}+4\alpha_{qc}-\frac{2}{3}\alpha_{cc}-\frac{8\beta}{3m_q^2} -\frac{16\beta}{m_qm_c} \nonumber\\
                  &\; -\frac{8\beta}{3m_c^2}+\frac{4\rho}{3m_q^2}-\frac{\gamma}{3m_q^2}.
\label{Eq:mass_qc_qc_1}
\end{align}
For $\cal{D}\bar{\cal{D}}$ states with $I^G J^{PC}=0^+2^{++}$ and $I^G J^{PC}=1^-2^{++}$, whose spin wave functions are defined in Eq.~(\ref{Eq:Spin_wave_qc_qc}), the masses are, respectively, given by
\begin{align}
M_{(11)2,0} = &\; 2M_{qc}-\frac{2}{3}\alpha_{qq}-4\alpha_{qc}-\frac{2}{3}\alpha_{cc}-\frac{8\beta}{3m_q^2} -\frac{16\beta}{m_qm_c} \nonumber\\
                   &\; -\frac{8\beta}{3m_c^2}-\frac{20\rho}{3m_q^2}-\frac{5\gamma}{3m_q^2},  \\
M_{(11)2,1} = &\; 2M_{qc}-\frac{2}{3}\alpha_{qq}-4\alpha_{qc}-\frac{2}{3}\alpha_{cc}-\frac{8\beta}{3m_q^2} -\frac{16\beta}{m_qm_c} \nonumber\\
                 &\; -\frac{8\beta}{3m_c^2}+\frac{4\rho}{3m_q^2}+\frac{\gamma}{3m_q^2}.
\label{Eq:mass_qc_qc_2}
\end{align}

The mass formulas for ${\cal{D}}_s{\bar{\cal{D}}}_{s}$ and ${\cal{D}}_s{\bar{\cal{D}}}$ $({\cal{D}}{\bar{\cal{D}}}_{s})$ can be calculated in the same way. To make the paper more concise, we exhibit them in Appendixes~\ref{Sect:sc_sc} and \ref{Sect:qc_sc}, respectively.

\subsection{Model parameters} \label{Sect:parameter}

Apart from the masses of $u(d)$, $s$, and $c$ quarks and the parameter $\beta$ in Eq.~(\ref{Eq:Hamiltonian_0}), whose values are taken from Refs.~\cite{Huang2018,Shi2019} as $m_{u(d)}=313$ MeV, $m_s=470$ MeV, $m_c=1650$ MeV, and $\beta=0.41$ fm$^{-3}$, the values of the following parameters in Eqs.~(\ref{Eq:Hamiltonian_0})$-$(\ref{Eq:Hamiltonian}) need to be determined before the calculation of the spectrum of hidden charm tetraquark states: $M_{qc}$, $M_{sc}$, $\alpha_{qq}$, $\alpha_{qs}$, $\alpha_{qc}$, $\alpha_{sc}$, $\alpha_{cc}$, $\alpha_{ss}$, $\rho$, and $\gamma$. 

Following Refs.~\cite{Maiani2005,Hogaasen2006,Hogaasen2006,Kim2016,Deng2015}, we attribute the  $X(3872)$, $Z_c(3900)$, and $\chi_{c2}(3930)$ as $[qc][\bar{q}\bar{c}]$ states with $I^GJ^{PC}=0^+1^{++}$, $1^+1^{+-}$, and $0^+2^{++}$, respectively, to fix the parameters $M_{qc}$, $\rho$, and $\gamma$. Following Refs.~\cite{JAFFE2005,Shi2019}, the heavy baryons $\Lambda_c$, $\Xi_c$, $\Omega_c^0$, $\Sigma_c$, $\Xi_c'$, $\Sigma_c^*$, and $\Xi^*_c$ are attributed to diquark-quark configurations and their masses are used to fix the parameters $\alpha_{qq}$, $\alpha_{qs}$, $\alpha_{qc}$, $\alpha_{sc}$, and $\alpha_{ss}$.

In Refs.~\cite{Wu2016,Maiani2016,Lebed2016}, the $X(4350)$ is treated as $[sc][\bar{s}\bar{c}]$ tetraquark state. But it is not quite known whether its quantum numbers $I^GJ^{PC}$ should be $0^+0^{++}$ or $0^+2^{++}$ due to the lack of experimental information. In the present work, we use the mass of $X(4350)$ to determine the parameter $M_{sc}$. In model I, we assume that the quantum numbers for $X(4350)$ are $I^GJ^{PC}=0^+0^{++}$, and in model II, we assume $I^GJ^{PC}=0^+2^{++}$ for $X(4350)$. The resulted values of $M_{sc}$ are $2252$ and $2205$ MeV, respectively, in models I and II.

The parameter $\alpha_{cc}$ can be fixed by the mass splits of $J/\psi$ and $\eta_c$, which gives
\begin{align}
\alpha_{cc}=-\frac{3}{32}\left(m_{J/ \psi}-m_{\eta_c}\right),
\end{align}
with $m_{J/\psi}$ and $m_{\eta_c}$ being the masses of $J/\psi$ and $\eta_c$, respectively.

All the model parameters needed in the present work are listed in Table~\ref{Tab:parameters}. 

\begin{table}[tb]
\caption{\label{Tab:parameters} Model parameters. The parameters $M_{qc}$, $M_{sc}$, $\alpha_{qq}$, $\alpha_{qs}$, $\alpha_{qc}$, $\alpha_{sc}$, $\alpha_{cc}$, and $\alpha_{ss}$ are in MeV. The parameters $\rho$ and $\gamma$ are in fm$^{-3}$. The quark masses and parameter $\beta$ are taken from Refs.~\cite{Huang2018,Shi2019} as $m_{u(d)}=313$ MeV, $m_s=470$ MeV, $m_c=1650$ MeV, and $\beta=0.41$ fm$^{-3}$.}
\renewcommand{\arraystretch}{1.2}
\begin{tabular*}{\columnwidth}{@{\extracolsep\fill}cccccc}
\hline\hline                                           
$\alpha_{qc}$  &   $\alpha_{sc}$ &  $\alpha_{cc}$  &  $\alpha_{qq}$ &   $\alpha_{qs}$  &  $\alpha_{ss}$   \\
 $-8.06$         &   $-8.98$             & $-10.59$      &  $-28.43$       &   $-21.70$        &   $-23.24$  \\  \hline
$M_{qc}$  & \multicolumn{2}{c}{$M_{sc}$} &  $\gamma$  &   $\rho$  \\
\cline{2-3}
              &   Model I   &   Model II  \\
$2059$    & $2252$  &  $2205$  & $0.03$        &   $0.11$   \\
\hline\hline
\end{tabular*}
\end{table}

\section{Numerical results}\label{Sect:Numerical_result}

With the values of model parameters listed in Table~\ref{Tab:parameters}, the masses of hidden charm tetraquark configurations $[qc][\bar{q}\bar{c}]$, $[sc][\bar{s}\bar{c}]$, and $[qc][\bar{s}\bar{c}]$ $\left([sc][\bar{q}\bar{c}]\right)$, which are denoted as $\cal{D}\bar{\cal{D}}$, ${\cal{D}}_s\bar{\cal{D}}_s$, ${\cal{D}}\bar{\cal{D}}_s$ (${\cal{D}}_s\bar{\cal{D}}$), respectively, in Sec.~\ref{Sect:wave function} and Sec.~\ref{Sect:Mass_formula}, can be obtained directly by use of the mass formulas given in Sec.~\ref{sec:mass_DDbar} and Appendixes~\ref{Sect:sc_sc} and \ref{Sect:qc_sc}.

\begin{table}[tb]
\caption{\label{mass_qcqc} Mass spectrum of tetraquark configurations $[qc][\bar{q}\bar{c}]$. The masses are in MeV. The first two columns show the quantum numbers and mass of each tetraquark configuration from our theoretical model, and the last three columns show the particle name, quantum numbers, and energy of each state advocated in PDG \cite{PDG2020}. In the third column, the states marked with ``$\ast$'' are input used to fix the model parameters, and the states marked with ``$?$'' are those that have more than one possible assignment.  }
\renewcommand{\arraystretch}{1.2}
\begin{tabular*}{\columnwidth}{@{\extracolsep\fill}lcccc}
\hline\hline                                           
 $I^GJ^{PC}$   &    $M$      &  Particle  & $\left(I^GJ^{PC}\right)_{\rm PDG}$  & $M_{\rm PDG}$   \\[3pt]   \hline
$0^+0^{++}$   &  $3795$   &      &      \\
$0^+0^{++}$   &  $3839$   & $\chi_{c0}(3860)$ & $0^+0^{++}$  &  $3862^{+26+40}_{-32-13}$  \\
$1^-0^{++}$     &  $3854$   &               &       \\
$1^-0^{++}$     &  $3910$   & $X(3940)?$              & $?^??^{??}$  & $3942^{+7}_{-6}\!\pm\!6$ \\
$0^+1^{++}$     &  $3869$   & $X(3872)^*$ & $0^+1^{++}$ & $3871.69\!\pm\!0.17$    \\
$1^-1^{++}$     &   $3934$   &  $X(3940)?$            & $?^??^{??}$  & $3942^{+7}_{-6}\!\pm\!6$    \\
$0^-1^{+-}$     &  $3830$   &               &      \\
$0^-1^{+-}$     &  $3868$   &       &                 \\
$1^+1^{+-}$      &  $3887$   & $Z_c(3900)^*$    & $1^+1^{+-}$   & $3888.4\!\pm\!2.5$       \\
$1^+1^{+-}$      &  $3942$   & $X(4020)$     & $1^+?^{?-}$  & $4024.1\!\pm\!1.9$        \\
$0^+2^{++}$     &  $3925$   & $\chi_{c2}(3930)^*$     & $0^+2^{++}$  & $3922\!\pm\!1$  \\
$1^-2^{++}$     &  $4000$    & $X(4050)$      & $1^-?^{?+}$   & $4051\!\pm\!14^{+20}_{-14}$       \\[2pt]
\hline\hline
\end{tabular*}
\end{table}

\begin{table}[tb]
\caption{\label{mass_scsc} Mass spectrum of tetraquark configurations $[sc][\bar{s}\bar{c}]$. The masses are in MeV. The first column shows the quantum numbers of each theoretical tetraquark configuration. The second and third columns show the masses calculated in model I and model II, where the $X(4350)$ is assumed to be a tetraquark state with $I^GJ^{PC} = 0^+0^{++}$ and $0^+2^{++}$, respectively. The last three columns show the particle name, quantum numbers, and energy of each state advocated in PDG \cite{PDG2020}. In the fourth column, the states marked with ``$\ast$'' and ``$?$'' are input used to fix the model parameters and have more than one possible assignment. Note that for $[sc][\bar{s}\bar{c}]$ configurations, $I=0$ and $G=C(-1)^I$.}
\renewcommand{\arraystretch}{1.2}
\begin{tabular*}{\columnwidth}{@{\extracolsep\fill}lccccc}
\hline\hline                                           
 $J^{PC}$    & $M_{\rm I}$  & $M_{\rm II}$  & Particle  & $\left(J^{PC}\right)_{\rm PDG}$ & $M_{\rm PDG}$ \\[3pt]                                               
\hline
$0^{++}$ & $4306$   &  $4212$    &            &        \\
$0^{++}$ & $4350$   &  $4257$   & $X(4350)^*?$   & $?^{?+}$  & $4350.6^{+4.6}_{-5.1}\!\pm\!0.7$   \\
$1^{++}$  & $4373$   &  $4280$   & $\chi_{c1} (4274)$ & $1^{++}$  & $4274^{+8}_{-6}$      \\
$1^{+-}$  & $4344$   &  $4251$   &           &         \\
$1^{+-}$  & $4383$  &  $4290$   &            &        \\
$2^{++}$ & $4443$  &  $4350$   & $X(4350)^*?$     & $?^{?+}$  & $4350.6^{+4.6}_{-5.1}\!\pm\!0.7$   \\[2pt]
\hline\hline
\end{tabular*}
\end{table}

\begin{table}[tb]
\caption{\label{mass_qcsc_bar} Mass spectrum of tetraquark configurations $[sc][\bar{q}\bar{c}]$ $\left([qc][\bar{s}\bar{c}]\right)$. The masses are in MeV. The first column shows the quantum numbers of each theoretical tetraquark configuration. The second and third columns show the masses calculated in model I and model II, respectively. The last three columns show the particle name, quantum numbers, and energy of each state reported in Ref.~\cite{Ablikim:2020hsk} for $Z_{cs}(3985)^-$ and in Ref.~\cite{Aaij:2021ivw} for $Z_{cs}(4000)^+$ and $Z_{cs}(4220)^+$. In the fourth column, the states marked with ``$?$'' are those that have more than one possible assignment.  }
\renewcommand{\arraystretch}{1.2}
\begin{tabular*}{\columnwidth}{@{\extracolsep\fill}lccccc}
\hline\hline                                           
 $J^{P}$ & $M_{\rm I}$ & $M_{\rm II}$ & Particle & $\left(J^{P}\right)_{\rm exp.}$ & $M_{\rm exp.}$ \\[3pt]        
\hline
$0^{+}$ & $4097$ & $4050$ & $Z_{cs}(3985)^- ?$ & $?^?$ & $3982.5^{+1.8}_{-2.6} \!\pm \!2.1$  \\
$0^{+}$ & $4141$ & $4095$ &        &    &  \\
$1^{+}$ & $4133$ & $4087$ & $Z_{cs}(3985)^- ?$  & $?^?$  & $3982.5^{+1.8}_{-2.6} \!\pm \!2.1$       \\
           &            &             & $Z_{cs}(4000)^+$  & $1^+$  & $4003\!\pm\! 6^{+4}_{-14}$       \\
$1^{+}$ & $4163$ & $4117$ &     &    &     \\
$1^{+}$ & $4173$ & $4126$ & $Z_{cs}(4220)^+$ & $1^?$ & $4216\!\pm\! 24^{+43}_{-30}$      \\
$2^{+}$ & $4231$ & $4185$  &     &    &   \\[2pt]  
\hline\hline
\end{tabular*}
\end{table}

The numerical results for the mass spectrum of $[qc][\bar{q}\bar{c}]$ are listed in Table~\ref{mass_qcqc}, where the first two columns show the quantum numbers $I^GJ^{PC}$ and the calculated mass of each tetraquark configuration, and the last three columns show the particle name, the quantum numbers $I^GJ^{PC}$, and the energy of each state advocated by the Particle Data Group (PDG) \cite{PDG2020}. In the third column, the states $X(3872)$, $Z_c(3900)$, and $\chi_{c2}(3930)$, each marked with an asterisk, are taken as input to fix the model parameters, as mentioned in Sec.~\ref{Sect:parameter}. The states marked with ``$?$'' are those that have more than one possible assignment. One sees from Table~\ref{mass_qcqc} that the $\chi_{c0}(3860)$ has a mass $M=3862^{+26+40}_{-32-13}$ MeV and quantum numbers $I^GJ^{PC}=0^+0^{++}$ in PDG \cite{PDG2020}. Our calculated mass of the tetraquark configuration $[qc][\bar{q}\bar{c}]$ with $I^GJ^{PC}=0^+0^{++}$ is 3839 MeV, very close to that of the experimentally observed $\chi_{c0}(3860)$, supporting the assignment of $\chi_{c0}(3860)$ as $[qc][\bar{q}\bar{c}]$ tetraquark state. The $X(4020)$ has a mass $M = 4024.1\!\pm\!1.9$ MeV and quantum numbers $I^GJ^{PC}=1^+?^{?-}$ in PDG \cite{PDG2020}. Our calculated mass of the $[qc][\bar{q}\bar{c}]$ state with $I^GJ^{PC}=1^+1^{+-}$ is $3942$ MeV,  not far away from that of the experimentally observed $X(4020)$. Accommodating the $X(4020)$ to a $[qc][\bar{q}\bar{c}]$ tetraquark state will suggest a spin $1$ and parity positive for this particle. The $X(4050)$ has a mass $M=4051\!\pm\!14^{+20}_{-14}$ and quantum numbers $1^-?^{?+}$ in PDG \cite{PDG2020}. Our calculated mass of the $[qc][\bar{q}\bar{c}]$ configuration state with $I^GJ^{PC}=1^-2^{++}$ is $4000$ MeV, close to that of the experimentally observed $X(4050)$, which supports the explanation of $X(4050)$ as a $[qc][\bar{q}\bar{c}]$ tetraquark state and suggests a spin $2$ and parity positive for this state. The $X(3940)$ has a mass $3942^{+7}_{-6}\!\pm\!6$ in PDG \cite{PDG2020}, but its quantum numbers are still not clear. Our calculated masses of the $[qc][\bar{q}\bar{c}]$ states with $I^GJ^{PC}=1^-0^{++}$ and $1^-1^{++}$ are $3910$ and $3934$ MeV, respectively, both very close to the experimental mass of $X(3940)$. This indicates that the $X(3940)$ could be explained as a $[qc][\bar{q}\bar{c}]$ tetraquark state and its quantum numbers might be $I^GJ^{PC}=1^-0^{++}$ or $1^-1^{++}$.

The theoretical results for the mass spectrum of $[sc][\bar{s}\bar{c}]$ are listed in Table~\ref{mass_scsc}. There, the first column shows the quantum numbers of each theoretical tetraquark configuration, the second and third columns show the masses calculated in model I and model II, where the $X(4350)$ is assumed to be a tetraquark state with $I^GJ^{PC} = 0^+0^{++}$ and $0^+2^{++}$, respectively. The last three columns show the particle name, quantum numbers, and energy of each state advocated in PDG \cite{PDG2020}. In the fourth column, the states marked with ``$\ast$'' and ``$?$'' are input used to fix the model parameters and have more than one possible assignment. Note that the quantum numbers $I^G$ are not shown in Table~\ref{mass_scsc}, as one always has $I=0$ and $G=C(-1)^I=C$ for the $[sc][\bar{s}\bar{c}]$ configurations. One sees from Table~\ref{mass_scsc} that in model I the calculated mass of the $[sc][\bar{s}\bar{c}]$ state with $J^{PC} = 1^{++}$ is $4373$ MeV, which is $99$ MeV higher than the mass of $\chi_{c1}(4274)$, a state advocated in PDG \cite{PDG2020} with the same quantum numbers. In model II, the calculated mass of this state is $4280$ MeV, which is very close to the experimental value of $\chi_{c1}(4274)$, supporting the explanation that the $\chi_{c1}(4274)$ is an $[sc][\bar{s}\bar{c}]$ tetraquark state. Note that although the $\chi_{c1}(4140)$ has the same quantum numbers as $\chi_{c1}(4274)$, our theoretical results do not support the explanation of the $\chi_{c1}(4140)$ as a tetraquark state, as the calculated masses of $[sc][\bar{s}\bar{c}]$ with $J^{PC} = 1^{++}$ in both model I and model II are too far away from the experimental mass of $\chi_{c1}(4140)$. In Ref.~\cite{Giron:2020qpb}, the $\chi_{c1}(4140)$ was assumed to be an $[sc][\bar{s}\bar{c}]$ state with $J^{PC}=1^{++}$ to fix the model parameters, and the $\chi_{c1}(4274)$ was considered as a charmonium state $\chi_{c1}(3P)$. More elaborate works need to be done to understand the nature of the $\chi_{c1}(4140)$ and $\chi_{c1}(4274)$ states.

The numerical results for the mass spectrum of tetraquark configurations $[sc][\bar{q}\bar{c}]$ $\left([qc][\bar{s}\bar{c}]\right)$ are listed in Table~\ref{mass_qcsc_bar}. There, the first column shows the quantum numbers of each theoretical tetraquark configuration. The second and third columns show the masses calculated in model I and model II, where the $X(4350)$ is assumed to be a tetraquark state with $I^GJ^{PC} = 0^+0^{++}$ and $0^+2^{++}$, respectively. The last three columns show the particle name, quantum numbers, and energy of each state reported in Ref.~\cite{Ablikim:2020hsk} for $Z_{cs}(3985)^-$ and in Ref.~\cite{Aaij:2021ivw} for $Z_{cs}(4000)^+$ and $Z_{cs}(4220)^+$. In the fourth column, the states marked with ``$?$'' are those that have more than one possible assignment. The $Z_{cs}(3985)^-$ has $M = 3982.5^{+1.8}_{-2.6} \pm 2.1$ MeV \cite{Ablikim:2020hsk}. Our calculated masses of the $[sc][\bar{q}\bar{c}]$ states with $J^P=0^+$ and $1^+$ are $4097$ MeV and $4133$ MeV in Model I, and $4050$ MeV and $4097$ MeV in Model II, not far away from the experimentally observed value. The $Z_{cs}(4000)^+$ and $Z_{cs}(4220)^+$ have masses $M = 4003 \pm 6^{+4}_{-14}$ and $M = 4216\pm 24^{+43}_{-30}$ \cite{Aaij:2021ivw}, respectively, which are in line with our calculated masses of the $[qc][\bar{s}\bar{c}]$ with $J^P=1^+$. 

In Refs.~\cite{Lebed:2017min,Giron:2020fvd}, the $P$-wave tetraquark states of $[qc][\bar{q}\bar{c}]$ with $I^G J^{PC}=0^- 1^{--}$ were discussed with the model parameters being fixed by the experimental information of the $Y(4230)$, $Y(4260)$, $Y(4360)$, and $Y(4390)$ states. In the present work, we can, in principle, include the spin-orbit and orbit-orbit interactions in the Hamiltonian of Eq.~(\ref{Eq:Hamiltonian}) to analyze the possible $P$-wave tetraquark states. However, for the $[sc][\bar{s}\bar{c}]$ and $[qc][\bar{s}\bar{c}]$ $\left([sc][\bar{q}\bar{c}]\right)$ configurations, we do not have enough experimental information to fix the additional model parameters for spin-orbit and orbit-orbit interactions. Thus, we leave such an analysis to future work when more experimental information becomes available.

\section{Summary}\label{Sect:Summary}

In the present work, we employ a diquark model to explore the mass spectrum of the hidden charm tetraquark states composed of diquarks and antidiquarks. The effective model Hamiltonian we used contains explicitly the color-spin, spin, flavor-spin, and flavor dependent interactions. We systematically calculate the masses of all possible $[qc][\bar{q}\bar{c}]$, $[sc][\bar{s}\bar{c}]$, and $[qc][\bar{s}\bar{c}]$ $\left([sc][\bar{q}\bar{c}]\right)$ hidden charm tetraquark states, and compare them with those advocated by the PDG \cite{PDG2020}. 

Our results show that, apart from the $X(3872)$, $Z(3900)$, $\chi_{c2}(3930)$, and $X(4350)$ which are taken as input to fix the model parameters, the $\chi_{c0}(3860)$, $X(4020)$, $X(4050)$ can be explained as $[qc][\bar{q}\bar{c}]$ states with $I^GJ^{PC}=0^+0^{++}$, $1^+1^{+-}$, and $1^-2^{++}$, respectively, and the $\chi_{c1}(4274)$ can be explained as an $[sc][\bar{s}\bar{c}]$ state with $I^GJ^{PC}=0^+1^{++}$. In addition, the $X(3940)$ can be explained as a $[qc][\bar{q}\bar{c}]$ state with $I^GJ^{PC}=1^-0^{++}$ or $1^-1^{++}$, the $X(4350)$ can be explained as an $[sc][\bar{s}\bar{c}]$ state with $I^GJ^{PC}=0^+0^{++}$ or $0^+2^{++}$, the $Z_{cs}(3985)^-$ can be explained as an $[sc][\bar{q}\bar{c}]$ state with $J^{P}=0^{+}$ or $1^{+}$, and the $Z_{cs}(4000)^+$ and $Z_{cs}(4220)^+$ can be explained as $[qc][\bar{s}{c}]$ states with $J^P=1^+$. Our results do not support the explanation of the $\chi_{c1}(4140)$ which has $I^GJ^{PC}=0^+1^{++}$ as an $[sc][\bar{s}\bar{c}]$ tetraquark state. We also give the predictions of other possible $[qc][\bar{q}\bar{c}]$, $[sc][\bar{s}\bar{c}]$, and $[qc][\bar{s}\bar{c}]$ $\left([sc][\bar{q}\bar{c}]\right)$ hidden charm tetraquark states, and hope that these states could be searched for in experiments in the near future.

\begin{acknowledgments}
This work is partially supported by the National Natural Science Foundation of China under Grants No.~11475181, No.~11635009, and No.~12075018, the Fundamental Research Funds for the Central Universities, and the Key Research Program of Frontier Sciences of the Chinese Academy of Sciences under Grant No.~Y7292610K1.
\end{acknowledgments}

\appendix

\section{The spin matrix elements}  \label{Sect:spin_matrix}

As shown in Eqs.~(\ref{Eq:Spin_wave_qc_qc_0})$-$(\ref{Eq:Spin_wave_qc_sc}), the spin wave function for the $[q_1 q_2][\bar{q}_3\bar{q}_4]$ tetraquark state are denoted as $\Ket{S_{12}, S_{\bar{3}\bar{4}}}_{S_{12\bar{3}\bar{4}}}$, which is an abbreviation of $\Ket{(S_1,S_2)S_{12}, (S_{\bar{3}},S_{\bar{4}})S_{\bar{3}\bar{4}}}_{S_{12\bar{3}\bar{4}}}$, with the spin of each quark (antiquark) $S_1=S_2=S_{\bar{3}}=S_{\bar{4}}=1/2$. 

The spin operator ${\bm S}_i \cdot {\bm S}_j$ satisfies 
\begin{align}
{\bm S}_i \cdot {\bm S}_j  =&\; \frac{\left({\bm S}_i + {\bm S}_j\right)^2 - {\bm S}_i^2 - {\bm S}_j^2}{2} \nonumber \\
=&\; \frac{S_{ij}\left(S_{ij}+1\right) - S_i \left(S_i+1\right) - S_j \left(S_j+1\right)}{2} \nonumber \\
=&\; \frac{S_{ij}\left(S_{ij}+1\right)}{2} - \frac{3}{4}.
\end{align}

By using this relation, the matrix elements of the spin operators ${\bm S}_1 \cdot {\bm S}_2$ and ${\bm S}_{\bar{3}} \cdot {\bm S}_{\bar{4}}$ can be calculated straightforwardly, while for other spin operators, the recoupling of the spin wave functions via $9j$ symbols are needed to compute their matrix elements. Specifically, one has
\begin{multline}
\Braket{S_{12}, S_{\bar{3}\bar{4}} | {\bm S}_1 \cdot {\bm S}_2 | S'_{12}, S'_{\bar{3}\bar{4}}}_{S_{12\bar{3}\bar{4}}} \\
= \left[\frac{S_{12}\left(S_{12}+1\right)}{2} - \frac{3}{4}\right] \delta_{S_{12}S'_{12}} \delta_{S_{\bar{3}\bar{4}} S'_{\bar{3}\bar{4}}}, ~~\quad \label{Eq:Spin_matrix_12}
\end{multline}
\begin{multline}
\Braket{S_{12}, S_{\bar{3}\bar{4}} | {\bm S}_{\bar{3}} \cdot {\bm S}_{\bar{4}} | S'_{12}, S'_{\bar{3}\bar{4}}}_{S_{12\bar{3}\bar{4}}}\\
= \left[\frac{S_{\bar{3}\bar{4}}\left(S_{\bar{3}\bar{4}}+1\right)}{2} - \frac{3}{4}\right] \delta_{S_{12}S'_{12}}\delta_{S_{\bar{3}\bar{4}}S'_{\bar{3}\bar{4}}},  ~~\quad
\end{multline}
\begin{multline}
\Braket{S_{12}, S_{\bar{3}\bar{4}} | {\bm S}_1 \cdot {\bm S}_{\bar{3}} | S'_{12}, S'_{\bar{3}\bar{4}}}_{S_{12\bar{3}\bar{4}}}  \\
= \sum_{S_{1\bar{3}} S_{2\bar{4}}} \hat{S}_{12} \hat{S}_{\bar{3}\bar{4}} \hat{S}^2_{1\bar{3}} \hat{S}^2_{2\bar{4}} \hat{S}'_{12} \hat{S}'_{\bar{3}\bar{4}} \left[\frac{S_{1\bar{3}}\left(S_{1\bar{3}}+1\right)}{2} - \frac{3}{4}\right] \quad \\
\times \begin{Bmatrix}
  1/2  & 1/2  & S_{12}  \\  1/2  &  1/2  &  S_{\bar{3}\bar{4}}  \\  S_{1\bar{3}}  &  S_{2\bar{4}} & {S_{12\bar{3}\bar{4}}}
\end{Bmatrix}
\begin{Bmatrix}
  1/2  & 1/2  & S'_{12}  \\  1/2  &  1/2  &  S'_{\bar{3}\bar{4}}  \\  S_{1\bar{3}}  &  S_{2\bar{4}} & {S_{12\bar{3}\bar{4}}}
\end{Bmatrix},
\end{multline}
\begin{multline}
\Braket{S_{12}, S_{\bar{3}\bar{4}} | {\bm S}_2 \cdot {\bm S}_{\bar{4}} | S'_{12}, S'_{\bar{3}\bar{4}}}_{S_{12\bar{3}\bar{4}}} \\
= \sum_{S_{1\bar{3}} S_{2\bar{4}}} \hat{S}_{12} \hat{S}_{\bar{3}\bar{4}} \hat{S}^2_{1\bar{3}} \hat{S}^2_{2\bar{4}} \hat{S}'_{12} \hat{S}'_{\bar{3}\bar{4}} \left[\frac{S_{2\bar{4}}\left(S_{2\bar{4}}+1\right)}{2} - \frac{3}{4}\right]  \\
\times \begin{Bmatrix}
  1/2  & 1/2  & S_{12}  \\  1/2  &  1/2  &  S_{\bar{3}\bar{4}}  \\  S_{1\bar{3}}  &  S_{2\bar{4}} & {S_{12\bar{3}\bar{4}}}
\end{Bmatrix}
\begin{Bmatrix}
  1/2  & 1/2  & S'_{12}  \\  1/2  &  1/2  &  S'_{\bar{3}\bar{4}}  \\  S_{1\bar{3}}  &  S_{2\bar{4}} & {S_{12\bar{3}\bar{4}}}
\end{Bmatrix},
\end{multline}
\begin{multline}
\Braket{S_{12}, S_{\bar{3}\bar{4}} | {\bm S}_1 \cdot {\bm S}_{\bar{4}} | S'_{12}, S'_{\bar{3}\bar{4}}}_{S_{12\bar{3}\bar{4}}} \\
= \sum_{S_{1\bar{4}} S_{2\bar{3}}} (-1)^{S_{\bar{3}\bar{4}}+S'_{\bar{3}\bar{4}}} \hat{S}_{12} \hat{S}_{\bar{3}\bar{4}} \hat{S}^2_{1\bar{4}} \hat{S}^2_{2\bar{3}} \hat{S}'_{12} \hat{S}'_{\bar{3}\bar{4}} \qquad\qquad \\
\times \left[\frac{S_{1\bar{4}}\left(S_{1\bar{4}}+1\right)}{2} - \frac{3}{4}\right]  \qquad ~~ \qquad \\
\times \begin{Bmatrix}
  1/2  & 1/2  & S_{12}  \\  1/2  &  1/2  &  S_{\bar{3}\bar{4}}  \\  S_{1\bar{4}}  &  S_{2\bar{3}} & {S_{12\bar{3}\bar{4}}}
\end{Bmatrix}
\begin{Bmatrix}
  1/2  & 1/2  & S'_{12}  \\  1/2  &  1/2  &  S'_{\bar{3}\bar{4}}  \\  S_{1\bar{4}}  &  S_{2\bar{3}} & {S_{12\bar{3}\bar{4}}}
\end{Bmatrix},
\end{multline}
\begin{multline}
\Braket{S_{12}, S_{\bar{3}\bar{4}} | {\bm S}_2 \cdot {\bm S}_{\bar{3}} | S'_{12}, S'_{\bar{3}\bar{4}}}_{S_{12\bar{3}\bar{4}}} \\
= \sum_{S_{1\bar{4}} S_{2\bar{3}}} (-1)^{S_{\bar{3}\bar{4}}+S'_{\bar{3}\bar{4}}} \hat{S}_{12} \hat{S}_{\bar{3}\bar{4}} \hat{S}^2_{1\bar{4}} \hat{S}^2_{2\bar{3}} \hat{S}'_{12} \hat{S}'_{\bar{3}\bar{4}} \qquad\qquad \\
\times \left[\frac{S_{2\bar{3}}\left(S_{2\bar{3}}+1\right)}{2} - \frac{3}{4}\right] \qquad ~~ \qquad \\
\times \begin{Bmatrix}
  1/2  & 1/2  & S_{12}  \\  1/2  &  1/2  &  S_{\bar{3}\bar{4}}  \\  S_{1\bar{4}}  &  S_{2\bar{3}} & {S_{12\bar{3}\bar{4}}}
\end{Bmatrix}
\begin{Bmatrix}
  1/2  & 1/2  & S'_{12}  \\  1/2  &  1/2  &  S'_{\bar{3}\bar{4}}  \\  S_{1\bar{4}}  &  S_{2\bar{3}} & {S_{12\bar{3}\bar{4}}}
\end{Bmatrix},  \label{Eq:Spin_matrix_23}
\end{multline}
where we have used the symbol $\hat{S}_{ij} = \sqrt{2 S_{ij} + 1 }$ for conciseness.

By use of Eqs.~(\ref{Eq:Spin_matrix_12})$-$(\ref{Eq:Spin_matrix_23}), all the needed spin matrix elements can be calculated, and the results are listed in Table~\ref{Tab:Spin_matrix}.

\section{Mass formulas for ${\cal{D}}_s{\bar{\cal{D}}}_{s}$} \label{Sect:sc_sc}

For ${\cal D}_s{\bar{\cal D}}_s$ states with $J^{PC}=0^{++}$, whose spin wave functions are defined in Eq.~(\ref{Eq:Spin_wave_qc_qc_0}), the masses are determined by the following mass matrix:
\begin{align}
 \begin{pmatrix}
 M_{(00)0}    & M'_{(00-11)0} \\
 M'_{(00-11)0}  & M_{(11)0}                                                    
\end{pmatrix},
\label{Eq:mass_sc_sc_0}
\end{align}
where the subscripts of the diagonal matrix elements $M$ denote $\left(S_{{\cal{D}}_s}S_{\bar{\cal{D}}_s}\right)S$, and the subscripts of the transition matrix elements $M'$ denote $\left(S_{{\cal{D}}_s}S_{\bar{\cal{D}}_s} - S'_{{\cal{D}}_s}S'_{\bar{\cal{D}}_s}\right)S$, with $S$ being the total spin for the considered tetraquark states. Explicitly, one has
\begin{align}
M_{(00)0} = &\; 2M_{sc}+8\alpha_{sc}-\frac{8\beta}{3m_s^2}-\frac{16\beta}{m_sm_c}-\frac{8\beta}{3m_c^2}  -\frac{8\rho}{3m_s^2},  \\
M_{(11)0} = &\; 2M_{sc}+\frac{4}{3}\alpha_{ss}+\frac{4}{3}\alpha_{cc}-\frac{8\beta}{3m_s^2}-\frac{16\beta}{m_sm_c} -\frac{8\beta}{3m_c^2}  \nonumber \\
                &\; -\frac{8\rho}{3m_s^2}+\frac{4\gamma}{3m_s^2},  \\
M'_{(00-11)0} = &\; \frac{2}{\sqrt{3}}\alpha_{ss}-\frac{4}{\sqrt{3}}\alpha_{sc}+\frac{2}{\sqrt{3}}\alpha_{cc}+\frac{2\gamma}{\sqrt{3}m_s^2}.
\end{align}
Similarly, for ${\cal D}_s{\bar{\cal D}}_s$ states with $J^{PC}=1^{+-}$, whose spin wave functions are defined in Eq.~(\ref{Eq:Spin_wave_qc_qc_2}), the masses are determined by the following mass matrix:
\begin{align}
\begin{pmatrix}
 M_{(01)1}    & M'_{(01-11)1} \\
 M'_{(01-11)1}  & M_{(11)1}                                                    
\end{pmatrix},
\label{Eq:mass_sc_sc_1_negtive}
\end{align}
where 
\begin{align}
M_{(01)1} = &\; 2M_{sc}+\frac{2}{3}\alpha_{ss}+\frac{4}{3}\alpha_{sc}+\frac{2}{3}\alpha_{cc}-\frac{8\beta}{3m_s^2} -\frac{16\beta}{m_sm_c}  \nonumber\\
                &\; -\frac{8\beta}{3m_c^2}-\frac{8\rho}{3m_s^2}+\frac{2\gamma}{3m_s^2},  \\
M_{(11)1} = &\; 2M_{sc}+\frac{2}{3}\alpha_{ss}-\frac{4}{3}\alpha_{sc}+\frac{2}{3}\alpha_{cc}-\frac{8\beta}{3m_s^2} -\frac{16\beta}{m_sm_c} \nonumber\\
               &\; -\frac{8\beta}{3m_c^2}-\frac{8\rho}{3m_s^2}+\frac{2\gamma}{3m_s^2},  \\
M'_{(01-11)1} = &\; -\frac{4}{3}\alpha_{ss}+\frac{4}{3}\alpha_{cc}-\frac{4\gamma}{3m_s^2}.
\end{align}
For ${\cal{D}}_s{\bar{\cal{D}}}_{s}$ states with $J^{PC}=1^{++}$ and $J^{PC}=2^{++}$, whose spin wave functions are defined in Eqs.~(\ref{Eq:Spin_wave_qc_qc_1}) and (\ref{Eq:Spin_wave_qc_qc}), the masses are, respectively, given by
\begin{align}
M_{1,0,1} = &\; 2M_{sc}-\frac{2}{3}\alpha_{ss}+4\alpha_{sc}-\frac{2}{3}\alpha_{cc}-\frac{8\beta}{3m_s^2}-\frac{16\beta}{m_sm_c}  \nonumber\\
&\; -\frac{8\beta}{3m_c^2}-\frac{8\rho}{3m_s^2}+\frac{2\gamma}{3m_s^2}, \\
M_{1,1,2} = &\; 2M_{sc}-\frac{2}{3}\alpha_{ss}-4\alpha_{sc}-\frac{2}{3}\alpha_{cc}-\frac{8\beta}{3m_s^2}-\frac{16\beta}{m_sm_c}  \nonumber\\
&\; -\frac{8\beta}{3m_c^2}-\frac{8\rho}{3m_s^2}-\frac{2\gamma}{3m_s^2}.
\label{Eq:mass_sc_sc_2}
\end{align}

\section{Mass formulas for ${\cal{D}}{\bar{\cal{D}}}_{s}$ $({\cal{D}}_s{\bar{\cal{D}}})$ }  \label{Sect:qc_sc}

For ${\cal D}{\bar{\cal D}}_s$ (${\cal D}_s{\bar{\cal D}}$) states with $I J^P=\frac{1}{2} 0^{+}$, whose spin wave functions are defined in Eq.~(\ref{Eq:Spin_wave_qc_sc_0}), the masses are determined by the following mass matrix:
\begin{align}
 \begin{pmatrix}
 M_{(00)0}    & M'_{(00-11)0} \\
 M'_{(00-11)0}  & M_{(11)0}                                                    
\end{pmatrix},
\label{Eq:mass_qc_sc_0}
\end{align}
where the subscripts of the diagonal matrix elements $M$ denote $\left(S_{{\cal{D}}}S_{\bar{\cal{D}}_s}\right)S$ or $\left(S_{{\cal{D}}_s}S_{\bar{\cal{D}}}\right)S$, and the subscripts of the transition matrix elements $M'$ denote $\left(S_{{\cal{D}}}S_{\bar{\cal{D}}_s} - S'_{{\cal{D}}}S'_{\bar{\cal{D}}_s}\right)S$ or $\left(S_{{\cal{D}}_s}S_{\bar{\cal{D}}} - S'_{{\cal{D}}_s}S'_{\bar{\cal{D}}}\right)S$, with $S$ being the total spin for the considered tetraquark states. Explicitly, one has
\begin{align}
M_{(00)0} = &\; M_{qc}+M_{sc}+4\alpha_{qc}+4\alpha_{sc}-\frac{8\beta}{3m_qm_c}-\frac{8\beta}{m_qm_c}\nonumber\\
                 &\; -\frac{8\beta}{m_sm_c}-\frac{8\beta}{3m_c^2}+\frac{4\rho}{3m_qm_s},    \\
M_{(11)0} = &\; M_{qc}+M_{sc}+\frac{4}{3}\alpha_{qs}+\frac{4}{3}\alpha_{cc}-\frac{8\beta}{3m_qm_s}-\frac{8\beta}{m_qm_c} \nonumber\\
                 &\; -\frac{8\beta}{m_sm_c}-\frac{8\beta}{3m_c^2} + \frac{4\rho}{3m_qm_s}-\frac{2\gamma}{3m_qm_s},  \\
M'_{(00-11)0} = &\; \frac{2}{\sqrt{3}} \left(\alpha_{qs}-\alpha_{qc}-\alpha_{sc}+\alpha_{cc}\right)-\frac{\gamma}{\sqrt{3}m_qm_s}.
\end{align}
Similarly, for ${\cal D}{\bar{\cal D}}_s$ (${\cal D}_s{\bar{\cal D}}$) states with $I J^P=\frac{1}{2} 1^{+}$, whose spin wave functions are defined in Eq.~(\ref{Eq:Spin_wave_qc_sc_1}), the masses are determined by the following mass matrix:
\begin{align}
 \begin{pmatrix}
 M_{(10)1}    & M'_{(10-01)1}  & M'_{(10-11)1}\\
 M'_{(10-01)1}  & M_{(01)1}    & M'_{(01-11)1}\\   
  M'_{(10-11)1}  & M'_{(01-11)1}    & M_{(11)1}                                                
\end{pmatrix},
\label{Eq:mass_qc_sc_1}
\end{align}
where 
\begin{align}
M_{(10)1} = &\; M_{qc}+M_{sc}+4\alpha_{sc}-\frac{4}{3}\alpha_{qc}-\frac{8\beta}{3m_qm_s}-\frac{8\beta}{m_qm_c}   \nonumber\\
                &\; -\frac{8\beta}{m_sm_c}-\frac{8\beta}{3m_c^2}+\frac{4\rho}{3m_qm_s},   \\
M_{(01)1} = &\; M_{qc}+M_{sc}-\frac{4}{3}\alpha_{sc}+4\alpha_{qc}-\frac{8\beta}{3m_qm_s}-\frac{8\beta}{m_qm_c}  \nonumber\\
                &\; -\frac{8\beta}{m_sm_c}-\frac{8\beta}{3m_c^2}+\frac{4\rho}{3m_qm_s},  \\
M_{(11)1} = &\; M_{qc} + M_{sc} +\frac{2}{3} \left(\alpha_{qs} -\alpha_{qc} -\alpha_{sc} + \alpha_{cc}\right)   \nonumber\\
                &\; -\frac{8\beta}{3m_qm_s} -\frac{8\beta}{m_qm_c} -\frac{8\beta}{m_sm_c}-\frac{8\beta}{3m_c^2}+\frac{4\rho}{3m_qm_s} \nonumber \\
                &\; -\frac{\gamma}{3m_qm_s},  \\
M'_{(10-01)1} = &\; -\frac{2}{3} \left( \alpha_{qs} - \alpha_{qc} - \alpha_{sc} + \alpha_{cc} \right)+\frac{\gamma}{3m_qm_s},  \\
M'_{(10-11)1} = &\; -\frac{2\sqrt{2}}{3} \left( \alpha_{qs} - \alpha_{qc} + \alpha_{sc} - \alpha_{cc} \right) +\frac{\sqrt{2}\gamma}{3m_qm_s},   \\
M'_{(01-11)1} = &\; \frac{2\sqrt{2}}{3} \left( \alpha_{qs} + \alpha_{qc} - \alpha_{sc} - \alpha_{cc} \right) -\frac{\sqrt{2}\gamma}{3m_qm_s}.
\end{align}
For ${\cal D}{\bar{\cal D}}_s$ (${\cal D}_s{\bar{\cal D}}$) states with $I J^{P}=\frac{1}{2} 2^{+}$, whose spin wave functions are defined in Eq.~(\ref{Eq:Spin_wave_qc_sc}), the masses are given by
\begin{align}
M_{(11)2} = &\; M_{qc}+M_{sc}-\frac{2}{3}\alpha_{qs}-2\alpha_{qc}-2\alpha_{sc}-\frac{2}{3}\alpha_{cc} \nonumber\\
                &\; -\frac{8\beta}{3m_qm_s}  -\frac{8\beta}{m_qm_c}-\frac{8\beta}{m_sm_c}-\frac{8\beta}{3m_c^2}+\frac{4\rho}{3m_qm_s} \nonumber \\
                &\; +\frac{\gamma}{3m_qm_s}.
\label{Eq:mass_qc_sc_2}
\end{align}

\end{document}